\documentclass[journal]{IEEEtran}
\ifCLASSINFOpdf
\else
\fi

\usepackage{epsfig}
\usepackage{epstopdf}

\usepackage{amsmath}
\usepackage{amsfonts}

\usepackage{graphicx}
\usepackage{hyperref}
\usepackage{subcaption}
\usepackage{caption}

\usepackage{pgfplots}

\usepackage{algorithmic}
\usepackage{algorithm}

\begin{document}
%
\title{Towards Power Efficient DNN Accelerator Design on Reconfigurable Platform}

\author{\IEEEauthorblockN{Rourab Paul$^1$, Sreetama Sarkar$^2$, Suman Sau$^3$, Sanghamitra Roy$^4$, Koushik Chakraborty$^5$, Amlan~Chakrabarti$^6$}
\IEEEauthorblockA{Computer Science \& Engineering, Siksha O Anusandhan, India$^1$, \\ Electrical and Computer Engineering Technical University Munich, Germany$^2$,\\ Computer Science  \& Information Technology,  Siksha 'O' Anusandhan, India$^3$,\\ Dept. Electrical and Computer Engineering, Utah State University, Logan,  USA$^{4,5}$, \\School of IT, University of Calcutta, India$^6$\\
mail: rourabpaul@soa.ac.in}
}


%


\maketitle

\begin{abstract}
The exponential emergence of Field Programmable Gate Array (FPGA) has
accelerated the research of hardware implementation of Deep Neural Network
(DNN). Among all DNN processors, domain specific architectures, such as, Google's Tensor Processor Unit (TPU) have
outperformed conventional GPUs. However, implementation of TPUs in
reconfigurable hardware should emphasize energy savings to serve the green
computing requirement. Voltage scaling, a popular approach towards energy
savings, can be a bit critical in FPGA as it may cause timing failure if not done in
an appropriate way. In this work, we present an ultra low power FPGA implementation of a TPU for edge applications. We divide the systolic-array of a TPU 
into different FPGA partitions, where each partition uses different near threshold
(NTC) biasing voltages to run its FPGA cores. The biasing voltage for each
partition is roughly calculated by the proposed static schemes. However, further
calibration of biasing voltage is done by the proposed runtime scheme. Four
clustering algorithms based on the minimum slack value of different design paths of Multiply Accumulates (MACs) study the partitioning of FPGA. To overcome the timing failure caused by NTC, the  MACs which have higher
minimum slack are placed in lower voltage partitions and the MACs have lower minimum slack path are
placed in higher voltage partitions. The proposed architecture is simulated in a commercial platform : $Vivado$ with Xilinx $Artix-7$ FPGA and academic platform VTR with 22nm, 45nm, 130nm FPGAs. The simulation
results substantiate the implementation of voltage scaled TPU in FPGAs and also
justifies its power efficiency.

\end{abstract}

\section{Introduction}
The configurable logic block (CLB) and switch matrix of FPGAs are power-hungry, which makes FPGAs energy inefficient compared to ASICs. Recently many researchers \cite{cloudscale}, \cite{putnam} have reported CPU-FPGA based hybrid data center architectures which provide hardware acceleration facility for Deep Neural Networks (DNNs). Despite power inefficiency, FPGA becomes popular in the Cloud-Scale acceleration architecture due to its specialized hardware and the economic benefits of homogeneity. Therefore, reducing power in FPGA for DNN applications becomes a very relevant topic of research. 
B Salami et al. \cite{mutulu} has studied the timing failure vs biasing voltage of DNN implementation in FPGA. They have underscaled biasing voltage $V_{ccint}$ of the entire FPGA to increase the power efficiency of Convolutional Neural Network (CNN) accelerator by a factor of 3. A single $V_{ccint}$ for the entire FPGA might not be the most power efficient solution. Partitioning an FPGA according to the slacks and feeding different biasing voltages for different partitions can cause further reduction of power for CNN implementations.  In \cite{greentpu}, the authors have implemented a systolic array using near threshold (NTC) biasing voltage in ASIC, which can predict the timing failure of  multiplier-accumulators (MACs) placed inside the systolic array of TPU. The prediction of timing failure is based on $Razor$ flipflop \cite{razor}. Higher fluctuation of input bits increases the possibility of timing failure in NTC condition. In \cite{greentpu}, once the timing failure of a MAC is predicted by its internal $Razor$ flipflop, the biasing voltage of the MAC is boosted up. 
\par Targeting FPGA based DNN applications \cite{cloudscale}, our work investigates voltage scaling techniques of TPU in the FPGA platform. Different $V_{ccint}$ for each of the MACs in a systolic array will be an absurd implementation for FPGA, therefore this work partitions FPGA floor according to the minimum slack value of design paths of MACs. Each partition consists of a group of MACs having similar minimum slacks. Each partition is connected with different $V_{ccint}$. The proposed methodology abstracts the synthesis timing report from the $Vivado$ and $VTR$. tool. In a synthesized design, the $Vivado$  and $VTR$ timing engine estimate the net delays of paths based on connectivity and fanout. The clustering algorithms create clusters or groups based on the minimum slack of MACs. The clusters consist of MACs which have lower minimum slacks are placed in FPGA partitions with higher $V_{ccint}$ and the clusters of MACs which have higher minimum slacks are placed in FPGA partitions with lower  $V_{ccint}$. Here the $V_{ccint}$ provides power to a FPGA core. 
The tuning of $V_{ccint}$ with slack is done by unique $static-runtime$ strategy.
The circuit level challenges on the implementation of voltage scaling in FPGA platform are beyond the present scope of our article. However, the feasibility of implementing the necessary hardware for voltage scaling support is evident considering the successful implementations in other ASIC technologies. As  is unavailable in current FPGAs we have simulated the design for the validation of the claim. The contribution of the paper is as follows:
\begin{itemize}
\item This paper proposes a new CAD flow to create voltage scaled TPU in FPGA based platforms considering the trade off of circuit delay against biasing voltage.
\item The proposed algorithm divides the systolic array of TPU into different partitions. Each partition will have different $V_{ccint}$. The  $V_{ccint}$ in different partitions is scaled against the different minimum slacks of different MACs.
\item  The calibration of $V_{ccint}$ of different partitions is done by the proposed $runtime$ and $static$ schemes. 
\end{itemize}

The organization of the article is as follows: Sec. \ref{sec:flow} outlines our background of FPGA environment. The working principle of Razor flipflop to detect runtime timing failure is discussed in \ref{sec:razor}. The methodology of the proposed work is described in Sec. \ref{sec:contri}. Sec. \ref{sec:cluster} discusses the clustering algorithms. Result, implementation and conclusion are organized in Sec. \ref{sec:impl} and Sec. \ref{sec:con} respectively. 
\vspace{-5pt}
\section{Background: FPGA Environment}\label{sec:flow}
The proposed scheme has been simulated in the both commercial and academic CAD tools.
In our first approach, we have used the Xilinx $Vivado$ tool with Artix-7 FPGA. For the sake of more accurate power data, the proposed voltage scaled architecture is also simulated in VTR tool flow with 22nm, 45nm and 130nm academic FPGAs. 
\subsection{Vivado Environment}
A typical Xilinx FPGA in $Vivado$ environment has 3 conventional steps such as synthesis, implementation and bit file generation whereas the adopted tool flow of the proposed partitioned FPGA is divided into two environments: (i) $Vivado$ Environment for synthesis, implementation and bit file generation and (ii) Python Environment for clustering similar slacks. The entire tool flow is shown in Fig. \ref{fig:vivtoolflow}.
The $Vivado$ environment is involved with 3 sub-steps stated below:
\subsubsection{Synthesis}
\label{sec:synth}
$Vivado$ synthesis process transforms register transistor logic (RTL) to gate level representation. The synthesis process generates delays of all possible paths of the design. The timing report of the synthesis process contains 12 information such as name of the path, slack value, level, high fanout, path from, path to, total delay of path, logic delay, net delay, time requirement source clock and destination clock as shown in Table \ref{tab:synreport}. It is to be noted that the estimation of the slacks of each logic block is at a high level. The actual timing behavior of the design depends on the net delays after placement and routing. 
\begin{table*}[!htbp]
	\caption{A Fragment of Timing Report from Synthesis for 100 MgHz Clock}    
	\label{tab:synreport}
	\centering
	\resizebox{18.2cm}{!}{
\begin{tabular}{|c|c|c|c|c|c|c|c|c|c|c|c|}\hline
Name	&Slack	&Levels	&\shortstack{High\\ Fanout}	&From	&To	&\shortstack{Total\\ Delay}	&\shortstack{Logic\\ Delay}	&\shortstack{Net\\ Delay}	&\shortstack{Requir\\ement}	&\shortstack{Source\\ Clock}	&\shortstack{Destination\\ Clock}\\\hline
Path 1	&5.34	&8	&8	&GEN\_REG\_I[0].GEN\_REG\_J[1].uut/prev\_activ\_reg[1]/C	&GEN\_REG\_I[1].GEN\_REG\_J[1].uut/sig\_mac\_out\_reg[16]/D	&4.37	&2.80	&1.57	&10.00	&clk	&clk\\\hline
Path 2	&5.49	&8	&8	&GEN\_REG\_I[0].GEN\_REG\_J[1].uut/prev\_activ\_reg[1]/C	&GEN\_REG\_I[1].GEN\_REG\_J[1].uut/sig\_mac\_out\_reg[15]/D	&4.40	&2.83	&1.57	&10.00	&clk	&clk\\\hline
Path 3	&5.52	&9	&8	&GEN\_REG\_I[0].GEN\_REG\_J[1].uut/prev\_activ\_reg[1]/C	&GEN\_REG\_I[1].GEN\_REG\_J[1].uut/sig\_mac\_out\_reg[14]/D	&4.36	&2.89	&1.47	&10.00	&clk	&clk\\\hline
Path 4	&5.59	&9	&8	&GEN\_REG\_I[0].GEN\_REG\_J[1].uut/prev\_activ\_reg[1]/C	&GEN\_REG\_I[1].GEN\_REG\_J[1].uut/sig\_mac\_out\_reg[13]/D	&4.30	&2.83	&1.47	&10.00	&clk	&clk\\\hline
Path 5	&5.78	&7	&8	&GEN\_REG\_I[0].GEN\_REG\_J[1].uut/prev\_activ\_reg[1]/C	&GEN\_REG\_I[1].GEN\_REG\_J[1].uut/sig\_mac\_out\_reg[12]/D	&4.10	&2.54	&1.57	&10.00	&clk	&clk\\\hline
Path 6	&5.83	&7	&8	&GEN\_REG\_I[0].GEN\_REG\_J[1].uut/prev\_activ\_reg[1]/C	&GEN\_REG\_I[1].GEN\_REG\_J[1].uut/sig\_mac\_out\_reg[11]/D	&4.05	&2.49	&1.57	&10.00	&clk	&clk\\\hline
		\end{tabular}
}
\end{table*}
\subsubsection{Implementation}
The $Vivado$ Implementation process is a timing-driven flow that transforms a logical netlist and constraints (Xilinx Design Constraints format) into a placed and routed design to make it ready for the bitstream generation process. In our proposed tool flow, the logical netlist is provided by the $Vivado$ synthesis process but the Xilinx Design Constraints (XDC) is generated by a python script. The clustered MACs are considered for placing in a specific location on the FPGA floor.
\subsubsection{Bit File Generation}
Once the placement and routing are completed by the implementation process, the flow generates bitstream of the systolic array. The Xilinx bitstream generation program produces a bitstream for the Xilinx device configuration. If there is any requirement of a processor, the design may include software application data.
\begin{figure}[!htb]
\centering
\vspace{-5pt}
\includegraphics[scale=0.38]{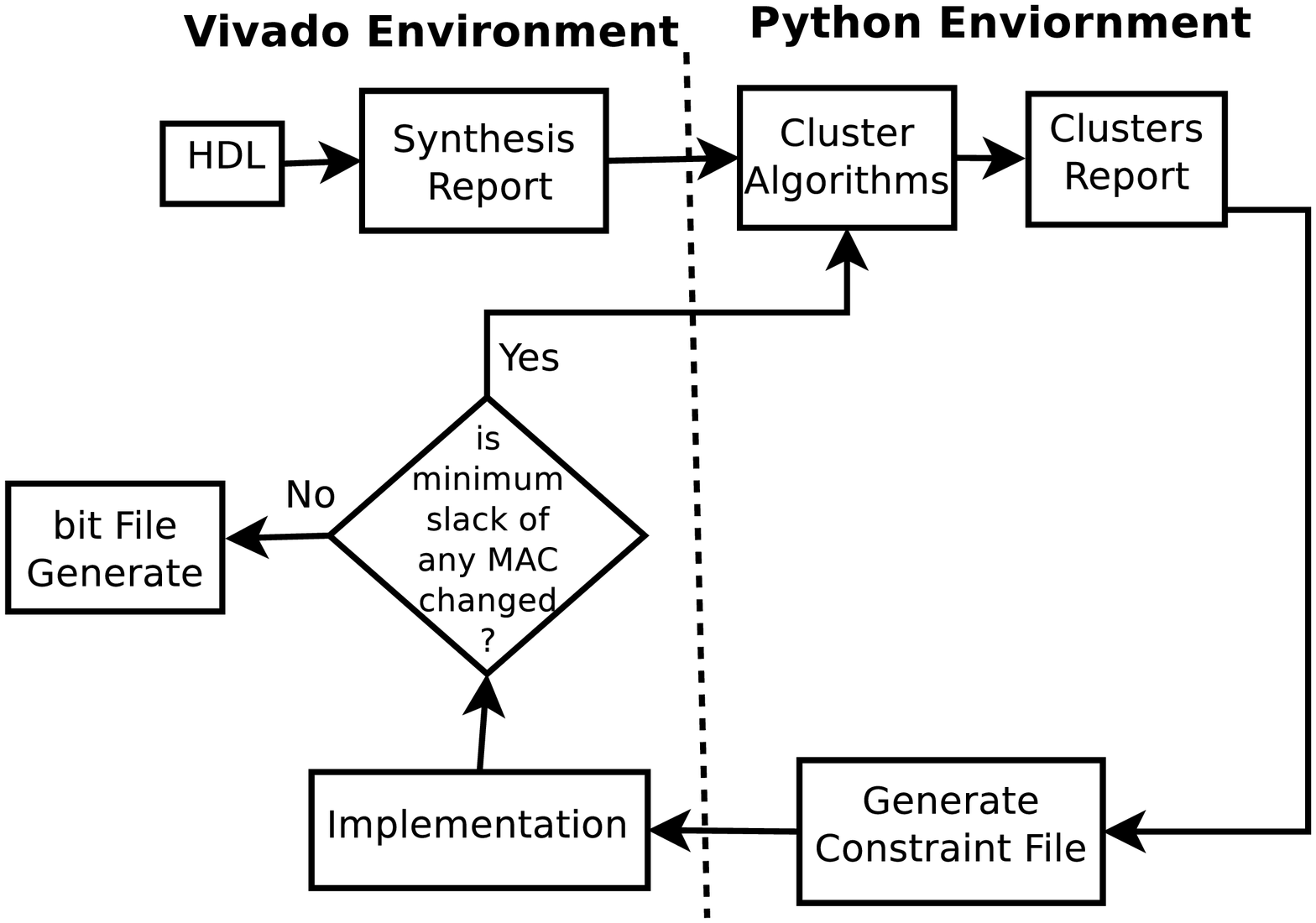}
\caption{Vivado Tool Flow}
\label{fig:vivtoolflow}
\end{figure}

\begin{figure*}[!htb]
\centering
\includegraphics[scale=1.0]{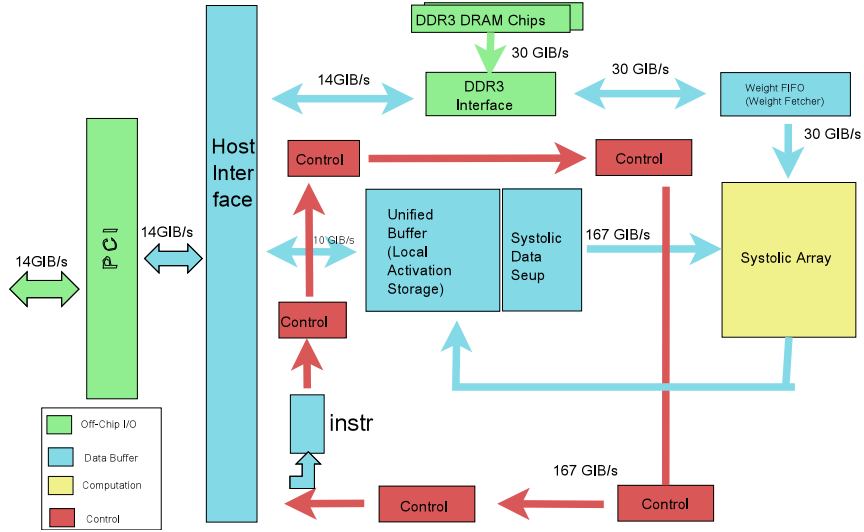}
\caption{TPU Architecture}
\label{fig:tpu}
\end{figure*}  
\subsection{VTR Environment}
In a commercial CAD environment biasing voltage is fixed.
The Verilog to Routing (VTR) \cite{vtr} tool is an open source academic CAD tool flow for FPGA architecture which allows voltage scaling technology. The VTR contains 3 separate tools such as Odin II \cite{odinii}, ABC \cite{abc} and VPR \cite{vpr}.
\subsubsection{Synthesis}
The synthesis process of the proposed VTR tool flow is processed by Odin II and ABC. Odin II elaborates and synthesizes HDL into FPGA architectural primitives like FFs, multipliers, adders. Thereafter the circuit logic is handled by ABC to perform technology independent logic optimizations, then technology maps the soft logic to LUTs. The information in the timing report generated by ABC is similar to the Vivado synthesis report. The different slack value of different design paths in synthesis report is used in cluster algorithms. 
\subsubsection{Implementation}
The VPR \cite{vpr} tool is a part of VTR flow which is used for the physical implementation of the circuit on the target FPGA architecture along with Synopsys Design Constraints File (sdc). In VTR flow, the logical netlist is provided by Odin II and ABC synthesis process but the SDC is generated by a python script. The clustered slack values generated by the python script are considered for placing the logic paths in a specific location on the FPGA floor. At the end, VPR analyzes the circuit implementation to generate area, speed and power data, and a
post-implementation netlist. Many commercial CAD tools like Titan Flow uses Intel’s Quartus  and Yosys used VPR for logic synthesis, optimization and technology mapping.
\par
\begin{figure*}[!htb]
\centering
\vspace{-5pt}
\includegraphics[scale=0.46]{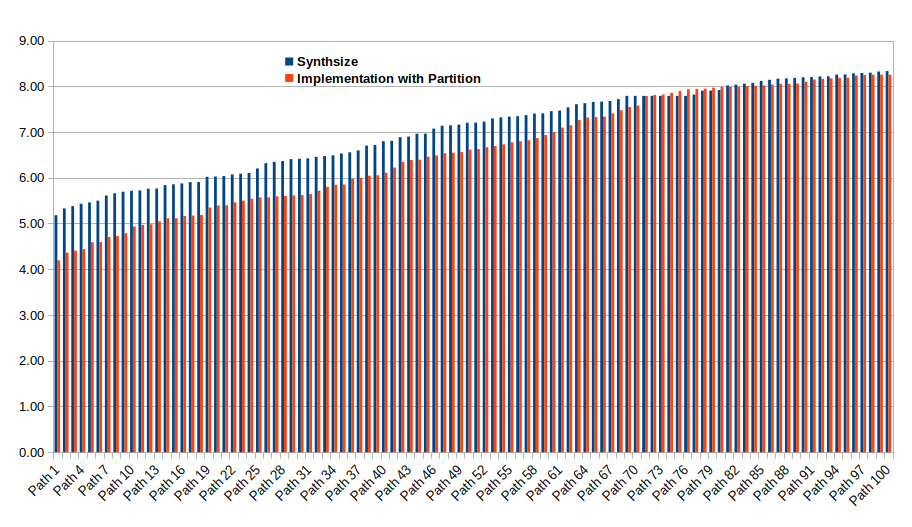}
\caption{100 worst Set up paths in Vivado}
\vspace{-5pt}
\label{fig:setup}
\end{figure*}

\begin{figure*}[!htb]
\centering
\vspace{-5pt}
\includegraphics[scale=0.49]{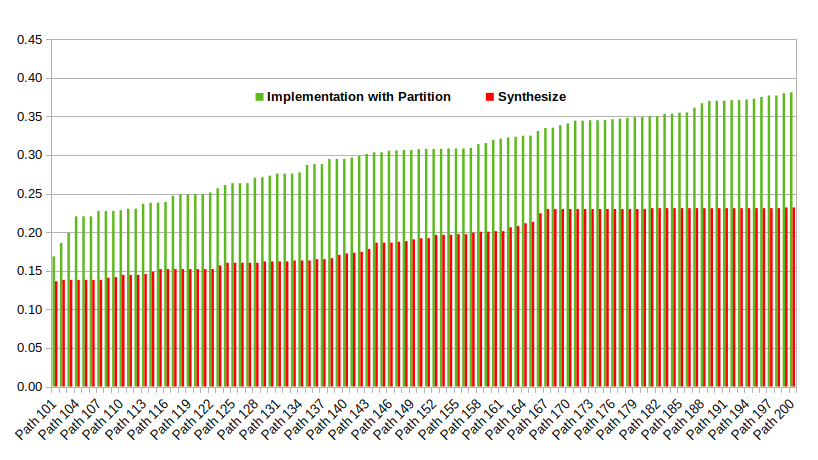}
\caption{100 worst Hold paths in Vivado}
\vspace{-5pt}
\label{fig:hold}
\end{figure*}

 There is a possibility that after the partitioning of the systolic array, delays of design paths from the implementation process may differ from delays of design paths from the synthesis process. If the minimum slack of any MAC is changed due to this partitioning, the entire design needs to re-cluster based on the new minimum slacks of MACs. Fig. \ref{fig:setup} and Fig. \ref{fig:hold} report the differecnes of delays of 100 worst design paths of synthesis process and implementation (after partition) process. Fig. \ref{fig:setup} and Fig. \ref{fig:hold} shows that partitioning process does not effect design paths significantly. The little proportional changes in the delays of design paths could not effect the minimum slack of MACs in each partiontion. Therefore, partitioning based on minimum slack of MACs in each partiontion will effective    
\begin{figure}[!htb]
\centering
\vspace{-5pt}
\includegraphics[scale=0.33]{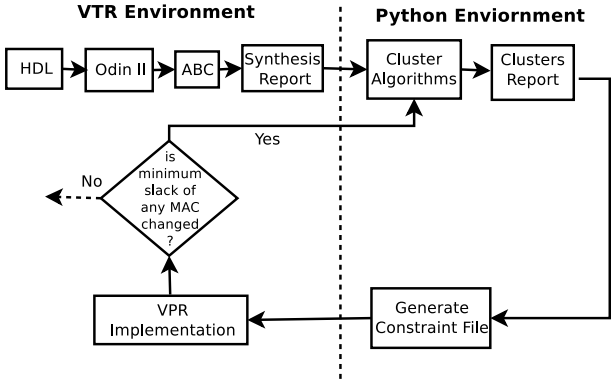}
\caption{VTR Tool Flow}
\vspace{-5pt}
\label{fig:VTRtoolflow}
\end{figure}

\subsection{Python Environment}
The contribution of the paper lies in augmenting the standard FPGA design tool flow by incorporating a python-based environment, which consists of a script to run three subsequent processes such as choice of $Clustering~Algorithms$, $Cluster~Generation$ and $Constraint~Generation$.
\subsubsection{Choice of Clustering Algorithms} A clustering algorithm suited to the requirements is chosen at this step. As stated in Sec. \ref{sec:contri}, this paper investigates 4 commonly-used clustering algorithms such as Hierarchical, K-means, Mean-shift and DBSCAN. 
 \subsubsection{Cluster Generation}
 We have assumed that the FPGA is divided into a few partitions and each partition has a different biasing voltage $V_{ccint}$.
 The clustering algorithms create few groups of MACs. The MACs having similar minimum slacks form a group and they are placed in the same FPGA partition.
 \subsubsection{Constraint Generation}
 Xilinx uses a constraint file format (XDC) to specify the coordinates of different paths of the proposed systolic array. The XDC file is generated by the python script.
 \subsection{Clustering MACs based on their Minimum Slacks}
The idea of voltage scaling for partitioned systolic array was initially based on the slacks generated from the synthesis report. The slack based clustering can group different similar design paths belonging to different MACs which may be placed in the different physical locations of FPGA floor by the placement and routing algorithm. For the slack based design path partitioning approach, intervention of the proposed tool's script is far more as compared to the existing placement and routing process of existing EDA tools. As a result, the timing parameters reported by the synthesis process are varied significantly after the placement and routing process of existing EDA tools at implementation level. For  4 partitions, 16 $\times$ 16 systolic array, the Vivado tool generates 6.23 ns critical path. The same design gives 11.93 ns critical path after placement routing which is almost two times the critical path generated from the synthesis report. We have noticed that the placement and routing process of slack based partitioning of 64 $\times$ 64 systolic array takes 10 to 14 hours in i5, 8GB Linux platform. Later, instead of clustering design paths based on slack, clustering is performed on MACs using their minimum slack values. We find clustering MACs based on their minimum slack is reasonable and better compare to the previous one for the below reasons:

 \begin{itemize}
 \item For the clustering of MACs based on their minimum slack, the intervention of vendor's technology dependent placement and routing algorithm is far more compared to the previous idea. As a result, the critical path variation in synthesis and implementation process is very minimum. 
 \item Placing all design paths in constraints file is much complicated compare to placing entire MACs in constraint file. 
 \item The routing of wires on the FPGA floor is comparatively simpler for the MAC clustering based on their minimum slacks.
 \end{itemize}

\subsection{Razor Flipflop}
\label{sec:razor}
\begin{figure}[!htb]
\centering
\includegraphics[scale=0.45]{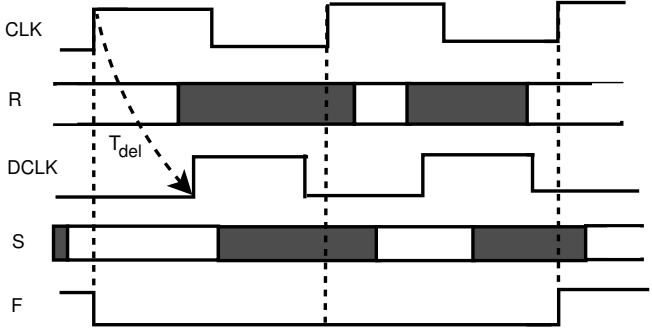}
\vspace{-5pt}
\caption{Timing Diagram of Fault Detection}
\vspace{-10pt}
\label{fig:tdr}
\end{figure}

Razor flipflop can be implemented in FPGA \cite{razor} by inserting a shadow flipflop which is running by a delayed clock. We have assumed that a circuit register $R$ is lying at the end of one or more timing paths originating from any of the source registers. The shadow register $S$ samples the same data as R but on a delayed clock $DCLK$ which is lagged by $T_{del}$ from the main clock $CLK$. Any data that arrives after R samples but before S samples will cause a discrepancy between the two registers that is detected by the error flag $F$. In each MAC unit, this razor flipflop is placed. The multiplication and addition process in each MAC is computed using the rising edge of $CLK$ and $DLK$.  The $CLK$ driven output of the multiplication and addition process of each MAC is stored in $R$ register. The $DCLK$ driven output of multiplication and addition process of each MAC is stored in shadow register $S$. The inclusion of razor doubles the number of multiplier and adder required for the systolic array but it can detect runtime failure occured in MACs due to the near threshold biasing voltage. The timing diagram of the razor is shown in Fig. \ref{fig:tdr}.

\section{Hybrid Configuration: Static \& Runtime Schemes} \label{sec:contri}
To mitigate timing failure issue in the critical voltage region, we adopted two sequential schemes such as (i)Static scheme which is involved with FPGA partitioning and rough  $V_{ccint_i}$ estimation depending on the FPGA technology. (ii) Runtime scheme to calibrate suitable $V_{ccint_i}$ for each partition of FPGA using $Razor$ flipflop. Each partition of FPGA consists a group of MACs. All the groups of MACs form a systolic array of the TPU. Apart from systolic array, TPU has memory to store active and weight inputs, PCI interface, controlling circuitry etc. The architecture of TPU is shown in Fig. \ref{fig:tpu}.
\subsection{Static Scheme} \label{sec:offline}
The proposed staic scheme works on $Vivado$ and $Python$ environments. As shown in Fig. \ref{fig:vivtoolflow}, synthesis is the first step of the proposed tool flow, which takes a netlist of complex logic blocks (CLBs) of systolic array generated from $Vivado$ tool. This netlist from the synthesis report is generated after technology mapping and packing stages which contain time slacks of all the possible paths of the systolic array. The proposed approach considers only nodes along paths because \textbf{(i)} the nodes along the path have data dependencies, which should be placed in the same FPGA partition even without considering the voltage scaling \cite{dualvdd}. \textbf{(ii)} The slack values of the nodes along paths are usually close to each other.
The second step of the proposed methodology is involved with the choice of the clustering algorithm and cluster generation. 
As stated in Sec. \ref{sec:cluster}, the four clustering algorithms such as Hierarchy, K-Mean, Mean-Shift and DBSCAN create multiple cluster of MACs with the paths available in the synthesis report. Even for the same number of clusters, different algorithms classify the data-points slightly differently. 
\par The primary concern is to identify clusters of MACs, which can share the minimum slacks available across the other MACs. Even for the same number of clusters, different algorithms classify the data-points slightly differently. Unlike K-means algorithm, the Hierarchical, Mean-Shift and DBSCAN do not need the number of clusters to be specified beforehand. DBSCAN is found to perform the best in this case as it groups together data-points close by, has a reasonable time complexity and can also identify outliers. Hence, clustered paths returned by DBSCAN are chosen for subsequent simulations.
\par
\begin{figure}[!htb]
\centering
\vspace{-10pt}
\includegraphics[scale=0.31]{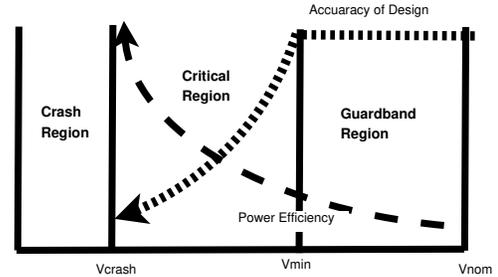}
\vspace{-5pt}
\caption{Voltage behaviour for $V_{ccint}$}

\label{fig:region}
\end{figure}

 Once the number of clusters is fixed we need to decide the voltage values of different FPGA partitions. In Fig. \ref{fig:region}., we illustrate 3 voltage regions in an FPGA, which is also supported by the research work in \cite{mutulu}. The voltage below FPGA crashing voltage $V_{crash}$ causes timing failure, which reduces the DNN accuracy near to zero. The region between minimum voltage $V_{min}$ and nominal voltage $V_{nom}$ is called guard band region where the DNN accuracy will be 100\% but power efficiency will be the least. In the critical region, the closer the voltage is to $V_{crash}$,
higher is the power efficiency and lower the DNN accuracy. Similarly, if $V_{ccint}$ is closer to $V_{min}$ in the critical region, the power efficiency decreases and DNN accuracy increases. In our proposed architecture, we assume the operating voltage range for the systolic array is $V_{crash}$ to $V_{min}$. If we have $n$ clusters computed by the chosen clustering algorithm we need $n$ partitions in FPGA. The primary $V_{ccint}$ estimation for each FPGA partition is computed by Algorithm \ref{alg:vs}. In Xilinx FPGA, the coordinates of circuit components are specified by two slice parameters $(X_i,Y_j)$. Each FPGA partition has range of these coordinates. The clustered MACs are placed in same FPGA partition by mentioning the slice parameters $(X_i,Y_j)$.
In the third step of the proposed methodology, each clustered path computed by the clustering algorithms is placed in a particular FPGA partition, which is restricted by specific $X_i,Y_j$ ranges. This restriction is done in the xdc file during $Generate~Constraint~File$ process. 
\par The rough $V_{ccint}$ calculation is done by Static Voltage Scaling algorithm shown in Algorithm \ref{alg:vs} which calculates a stepping voltage $V_s$ from $V_{min}$ and $V_{crash}$. Thereafter, the $V_{ccint_i}$ of the $i^{th}$ partition is calculated based on the stepping voltage $V_s$. The Static Voltage Scaling algorithm distributes $V_{ccint}$s for $i$ number of partitions for the range started from $V_{crash}$ to $V_{min}$.\\
\begin{minipage}{0.5\textwidth}
\begin{algorithm}[H]
\caption{Static Voltage Scaling} 
\label{alg:vs} 
\begin{algorithmic}[1]
\REQUIRE $V_{ccint}$, $V_{min}$, $V_{crash}$ \& $n$
\STATE $V_s$=$\frac{V_{min}-V_{crash}}{n}$
\STATE $V_l=V_{crash}$
\FOR{i=0 to n-1}
\STATE $V_{ccint_i}$ = $\frac {V_l+V_l+V_s}{2}$
\STATE $V_l=V_l+V_s$
\ENDFOR
\end{algorithmic}
\end{algorithm}
\end{minipage}
\begin{minipage}{0.5\textwidth}
\begin{algorithm}[H]
\caption{Runtime Voltage Scaling} 
\label{alg:on} 
\begin{algorithmic}[1]
\REQUIRE $V_{ccint}$, $V_{s}$
\FOR{i=0 to n-1}
 \IF{$timing\_fail-part-i==1$}
    \STATE $V_{ccint_i}$ = $V_{ccint_i}+V_s$
  \ELSE
    \STATE $V_{ccint_i}$ = $V_{ccint_i}-V_s$
  \ENDIF
\ENDFOR
\end{algorithmic}
\end{algorithm}
\end{minipage}
\subsection{Runtime Scheme}\label{sec:online}
The $V_{ccint_i}$ of the $i^{th}$ FPGA partition calculated by Algorithm \ref{alg:vs} is calibrated to $V_{ccint_i}$ pin of the $i^{th}$ FPGA partition. The calculation of $V_{ccint_i}$ by Algorithm \ref{alg:vs} is based on the number of partitions $n$ and the critical voltage region $V_{min}-V_{crash}$ which solely depends on the type of FPGA technology. However, the appropriate $V_{ccint_i}$ of the $i^{th}$ FPGA partition should also depend on the minimum slack values of MACs of that partition. At the static strategy we just calculate a rough estimation of $V_{ccint_i}$ where as  the runtime strategy calibrates $V_{ccint_i}$ according to the runtime timing failure of the systolic array. In the runtime scheme we used one of the most popular runtime timing error detection scheme, $Razor$, which uses double sampling flipflop to detect timing violation of pipeline stages. The $Razor$ flipflop is connected with every MACs of the systolic array to indicate its the timing failure. Each MAC has a timing failure flag which is controlled by the  $Razor$ flipflop. If any timing failure flag of any MAC placed in the  $i^{th}$ FPGA partition is high, the $V_{ccint_i}$ of that $i^{th}$ FPGA partition will be increased by one step. If all the timing failure flags of all MACs placed in the $i^{th}$ FPGA partition is low, the $V_{ccint_i}$ of that $i^{th}$ FPGA partition will be decreased by one step. Before starting the actual run of the proposed systolic array, if we have trial run, all the $V_{ccint_i}$ of all partitions will  be tuned accurately by this $runtime$ process. The voltage boosting circuit can be implemented  externally following the technique proposed in \cite{boost}.
\begin{figure}[!htb]
\includegraphics[scale=0.38]{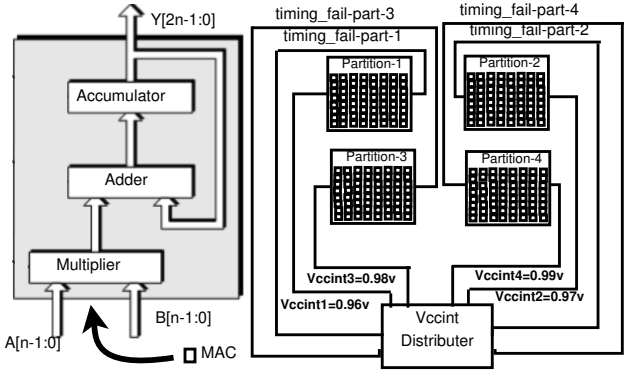}

\caption{Example : Partitioned FPGA, n=4}

\label{fig:part}
\end{figure}   
In Fig. \ref{fig:part}, we have shown that the cluster algorithm partitions the FPGA into 4 islands. The static scheme as stated in Sec. \ref{sec:offline} calculates 4 $V_{ccint_i}$ such as $V_{ccint_1}$,$V_{ccint_2}$, $V_{ccint_3}$ and $V_{ccint_4}$ for FPGA partition-1, partition-2, partition-3 and partition-4, respectively. The power distribution unit distributes $V_{ccint_i}$ such as $V_{ccint_1}$, $V_{ccint_2}$, $V_{ccint_3}$ and $V_{ccint_4}$ to FPGA partition-1, partition-2, partition-3 and partition-4 respectively. Thereafter, the TPU circuit can be on and the runtime scheme becomes functional. In Fig. \ref{fig:part}, 4 FPGA partitions, partition-1, partition-2, partition-3 and partition-4 have 4 flags  form $Razor$ flipflops, $timing\_fail-part-1$, $timing\_fail-part-2$, $timing\_fail-part-3$ and $timing\_fail-part-4$ respectively to detect the timing failure of the available partition of the FPGA. Each  $timing\_fail-part-i$ flag is ANDed value of all error detection flag all MACs placed in the $i^{th}$ partition. As shown in algorithm \ref{alg:on}, if $i^{th}$ timing failure flag from $i^{th}$ FPGA partitions becomes high, the power distribution network will step up  $V_{ccint_i}$ of that partition by $V_s$ else $V_{ccint_i}$ will be step down by $V_s$.

\begin{figure}[!htb]
\centering
\includegraphics[scale=0.37]{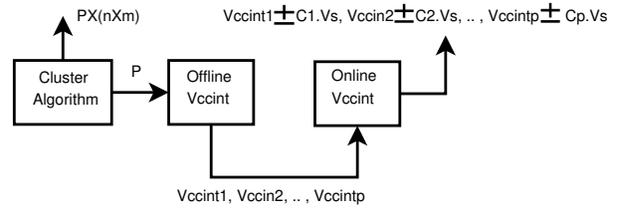}
\caption{Flow Diagram of Proposed Frame Work }
\label{fig:flow}
\end{figure} 
\section{Clustering Algorithms}\label{sec:cluster}
  We have investigated 4 clustering algorithms to group the MACs having similar minimum slacks. Algorithms can be chosen based on the design requirements: if we want to set a pre-defined number of clusters, or set hyperparameters to automatically determine the number of clusters. Different algorithms work well for different data distributions. Depending on our design requirements, we choose among the following four algorithms:

\subsection{Hierarchical}
\begin{figure}[!htb]
\centering
\vspace{-10pt}
\includegraphics[scale=0.33]{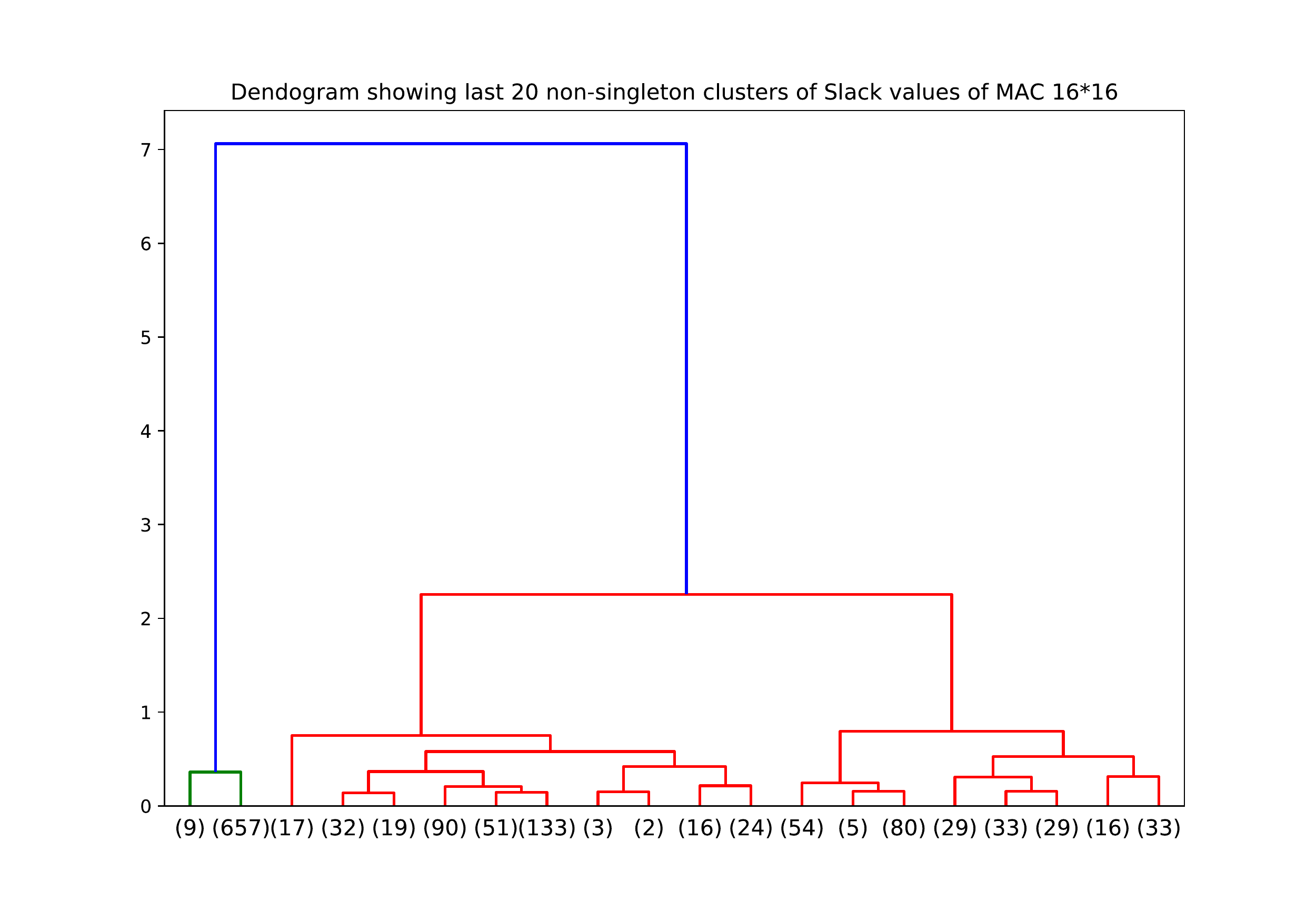}
\vspace{-5pt}
\caption{Dendrogram}
\vspace{-10pt}
\label{fig:dendrogram}
\end{figure}

\begin{figure*}[!htb]
\centering
\includegraphics[scale=0.112]{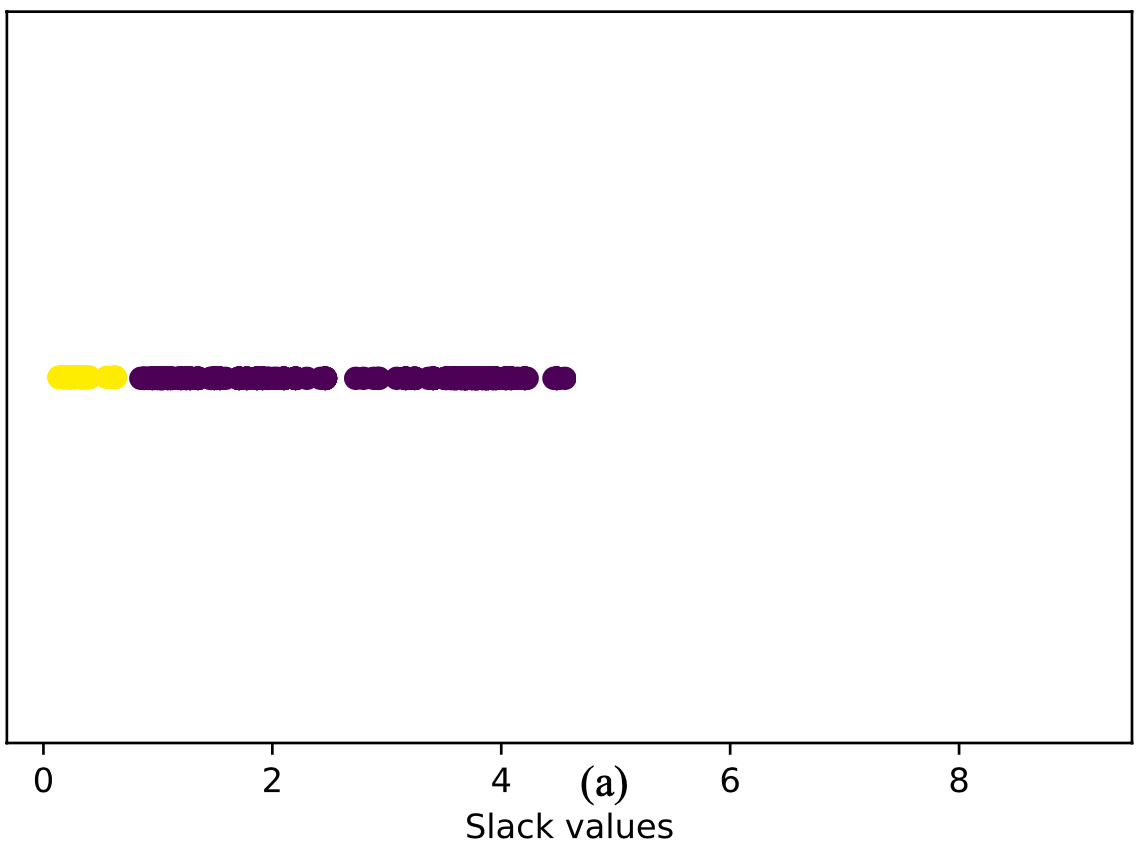}
\includegraphics[scale=0.112]{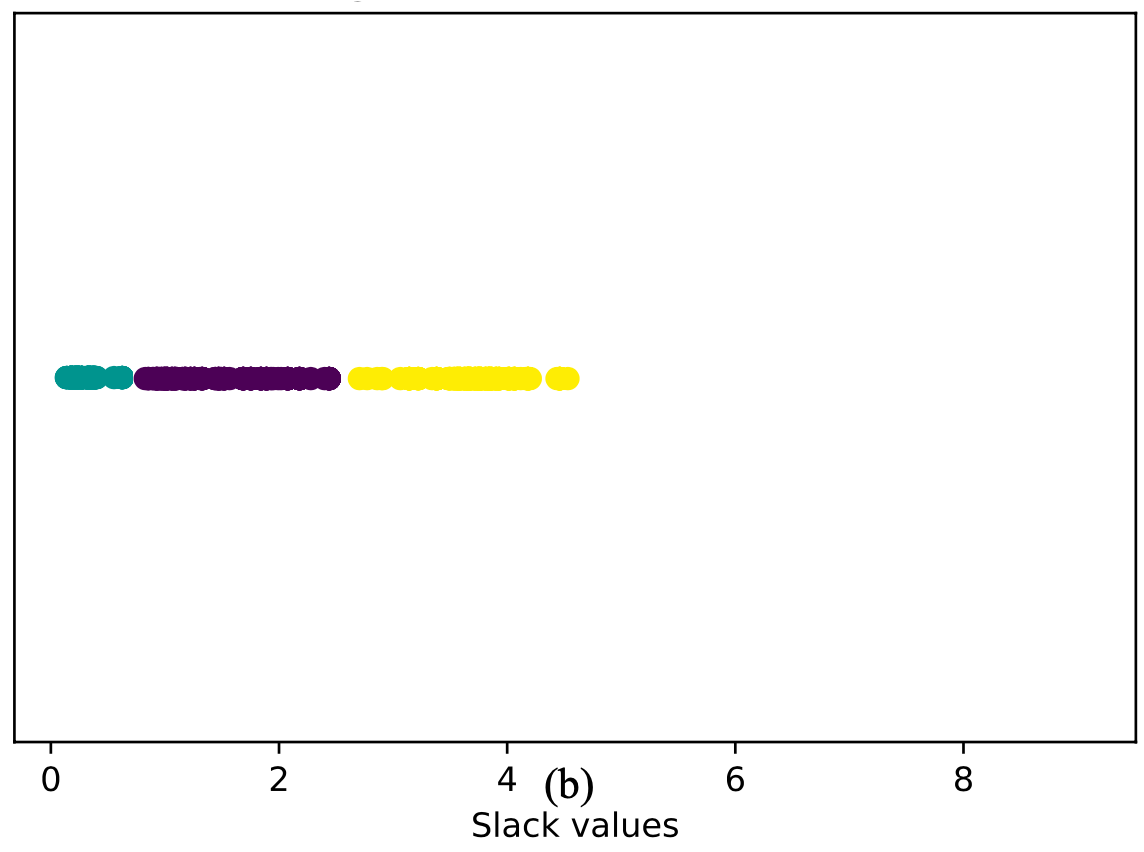}
\includegraphics[scale=0.114]{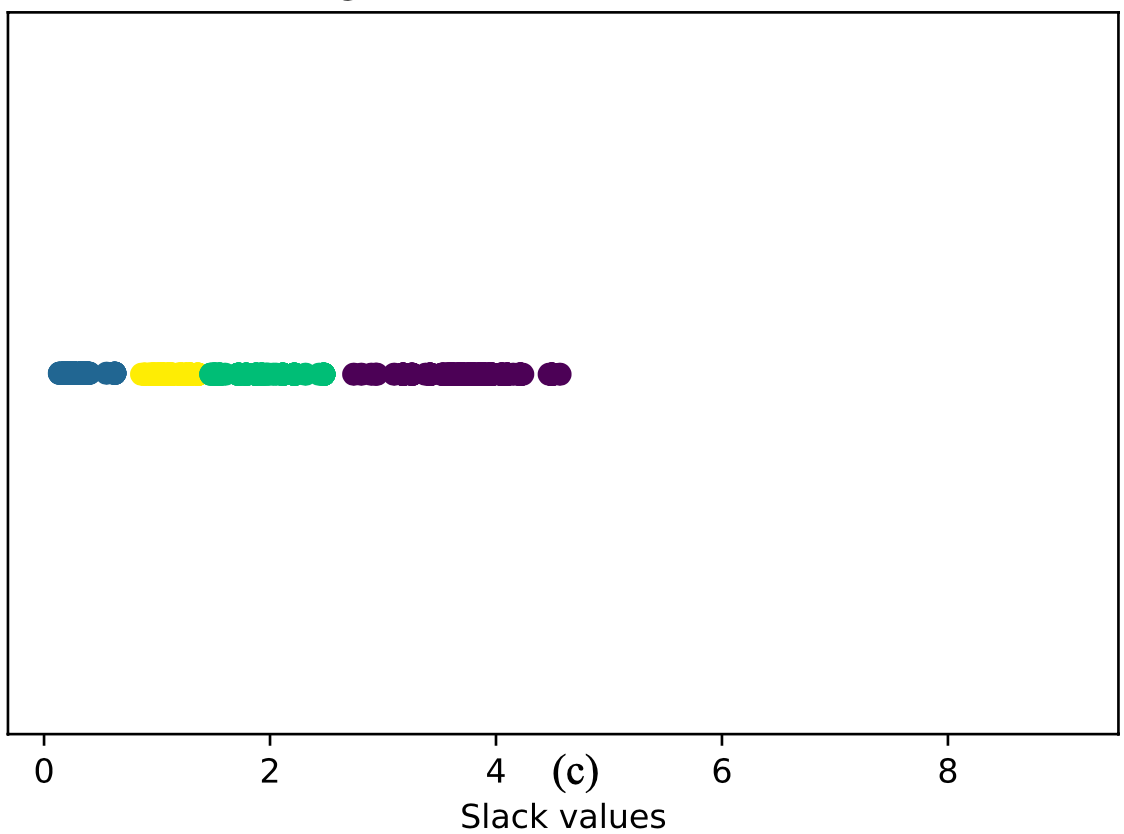}
\vspace{-5pt}
\caption{Hierarchical Cluster of Slack of Systolic Array $16\times 16$: a) \#clusters=2, b) \#clusters=3, c) \#clusters=4}

\label{fig:hier}
\end{figure*}

The hierarchical clustering \cite{hier} algorithm considers each data point as a single cluster and measures distance between two clusters based on a chosen distance measure (in this case, Euclidean distance).
The two clusters that are closest to each other are merged. The process is continued until all clusters have been merged into a single cluster (root of the dendrogram). As shown in fig. \ref{fig:dendrogram}, the dendrogram is a tree-like structure used for visualizing the hierarchy of clusters. 
The number of clusters can be decided from the dendrogram. The hierarchical algorithm is computationally expensive for large datasets, having a time complexity of $O(n^3)$ where $n$ is the number of data-points. As is evident from the dendrogram, the length of the branch joining the last two clusters is the highest, indicating they are the most dissimilar, followed by the third and fourth clusters. The result of classifying the slack values into 2, 3 and 4 clusters is illustrated in the fig. \ref{fig:hier}. Different clusters in Figs. \ref{fig:hier}, \ref{fig:kmean}, \ref{fig:mshift} and \ref{fig:dbscan} are indicated using different colours.

\subsection{K-Means Clustering}
\begin{figure*}[!htbp]
\centering
\vspace{-10pt}
\includegraphics[scale=0.112]{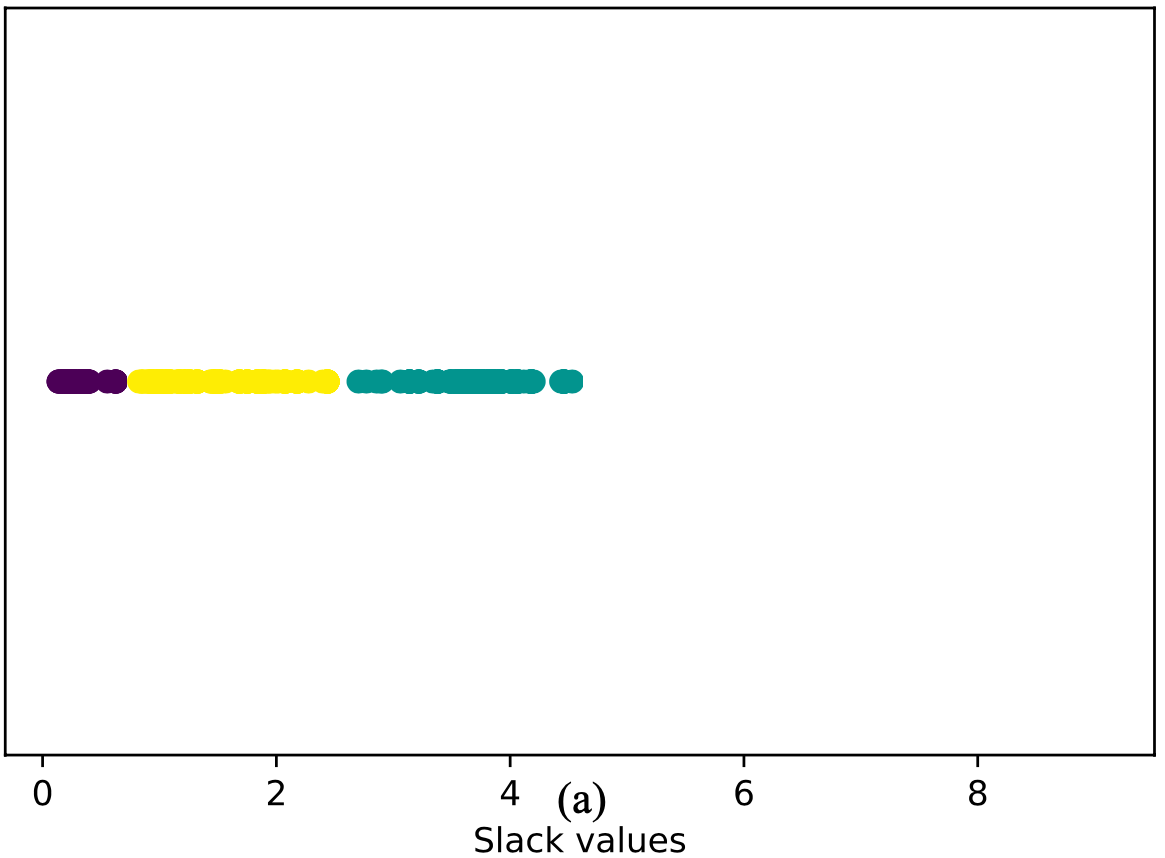}
\includegraphics[scale=0.112]{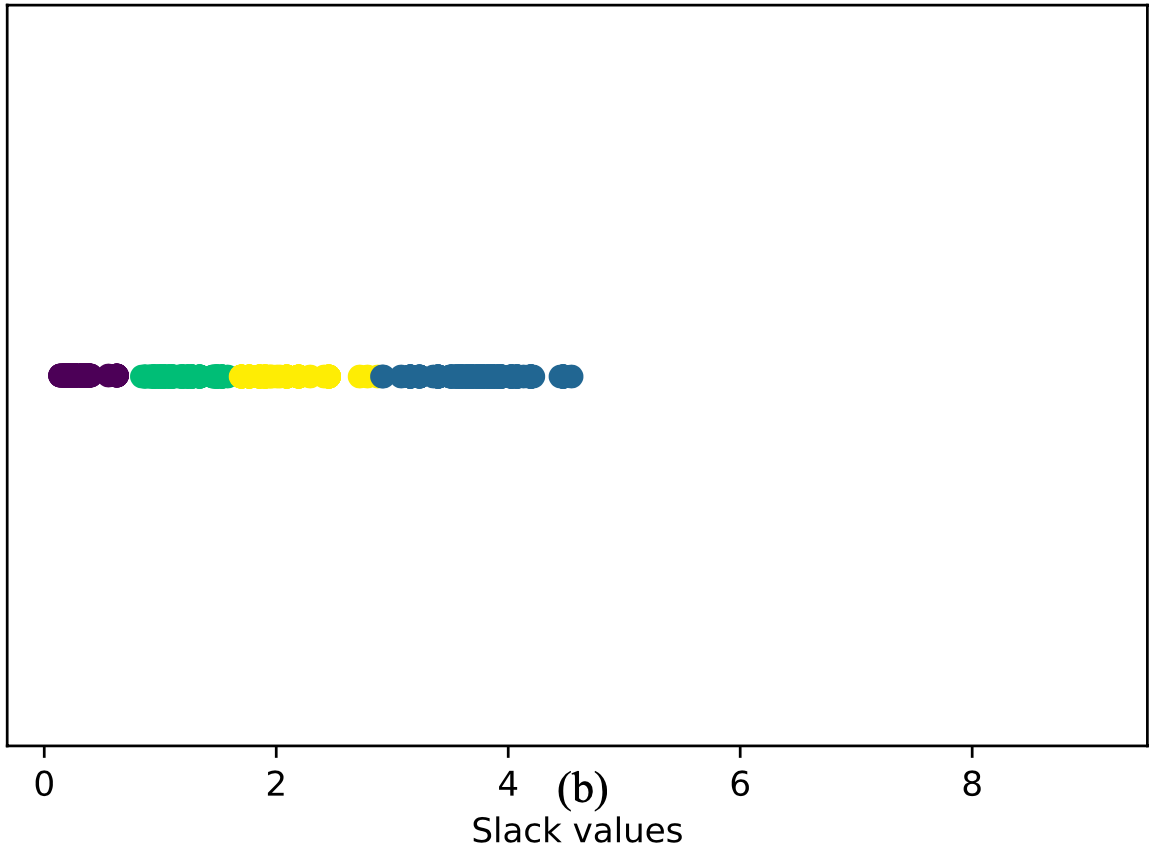}
\includegraphics[scale=0.114]{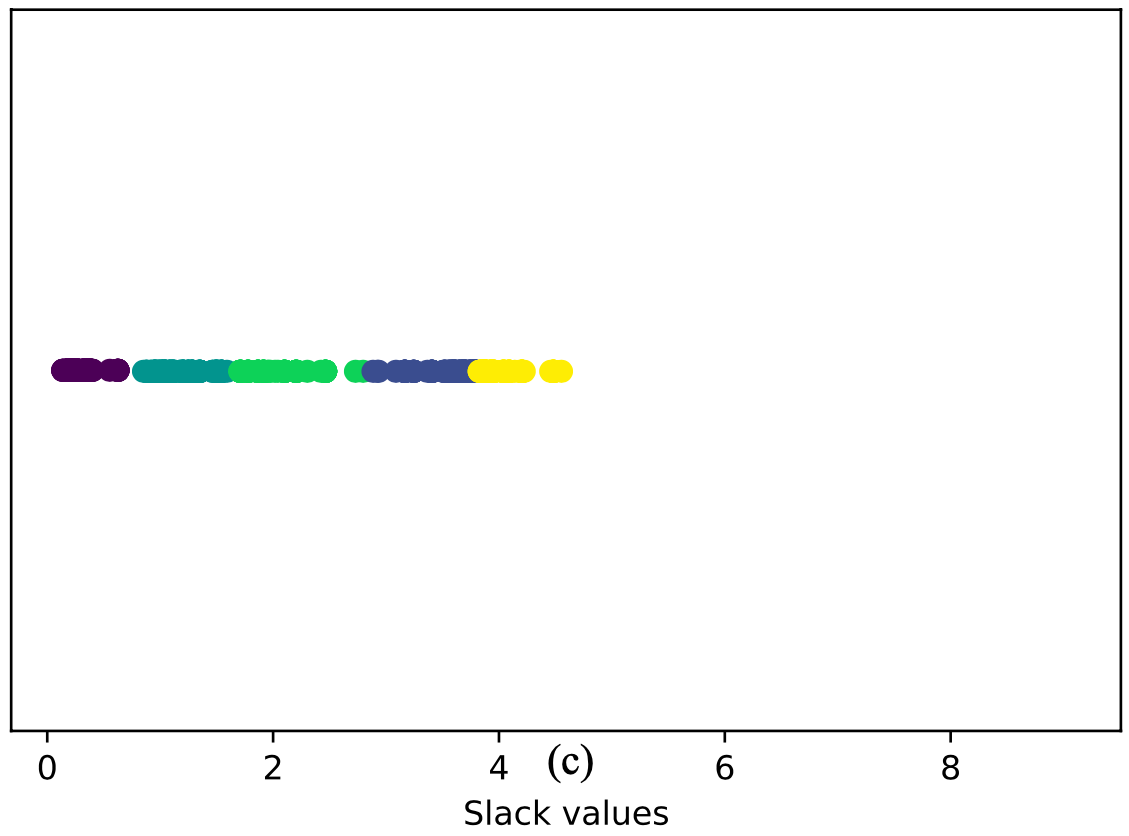}
\vspace{-5pt}
\caption{K Means Cluster of Slack of Systolic Array$16\times 16$: a) \#clusters=3, b) \#clusters=4, c) \#clusters=5}
\label{fig:kmean}
\end{figure*}

K-Means Clustering can cluster data into a predefined number of groups ($k$). At the beginning, $k$ cluster centers are randomly initialized \cite{kmean}. The algorithm computes the distance between each data-point and the cluster-centers and assigns data-points to the cluster whose center is closest to it. The cluster centers are then recomputed as the mean of the data-points belonging to that cluster. The process is repeated for a predefined number of steps or until cluster centers do not change significantly.
The K-Means Clustering is simple, fast, and its time complexity is $O(n)$. 
Fig. \ref{fig:kmean} illustrates the results of applying K-Means clustering algorithm on the minimum slack values of a $16\times16$ Systolic Array (256 MACs) for 3, 4 and 5 clusters. 

\subsection{Mean-Shift Clustering}
Mean Shift Clustering \cite{meanshift} is based on the idea of Kernel Density Estimation (KDE). KDE assumes that the data points are generated from an underlying distribution and tries to estimate the distribution by
assigning a kernel to each data point. The most commonly used kernel is the Gaussian or RBF kernel. 
The mean-shift algorithm is designed in a way that the points iteratively climb the KDE surface and are shifted to the nearest KDE peaks. It starts with a randomly selected point as the center of the RBF kernel. Thereafter, it proceeds by moving the kernel towards regions of higher density by shifting the center of the kernel to the mean of the points within the window (hence the algorithm is termed mean-shift). This is continued until shifting the kernel no longer includes more points. 
This algorithm does not need the number of clusters to be specified beforehand, but it is computationally expensive compared to K-Means (time complexity of $O(n*log(n))$ in lower dimension for sklearn implementation). The selection of the window size/radius $r$ can be non-trivial and plays a key-role in the success of the algorithm. 
Setting the radius as 0.4 for the slack values of a $16\times16$ Systolic array, yields 4 clusters as observed in the Fig. \ref{fig:mshift}.\\
\begin{minipage}{0.5\textwidth}
\begin{figure}[H]
\centering
\includegraphics[scale=0.17]{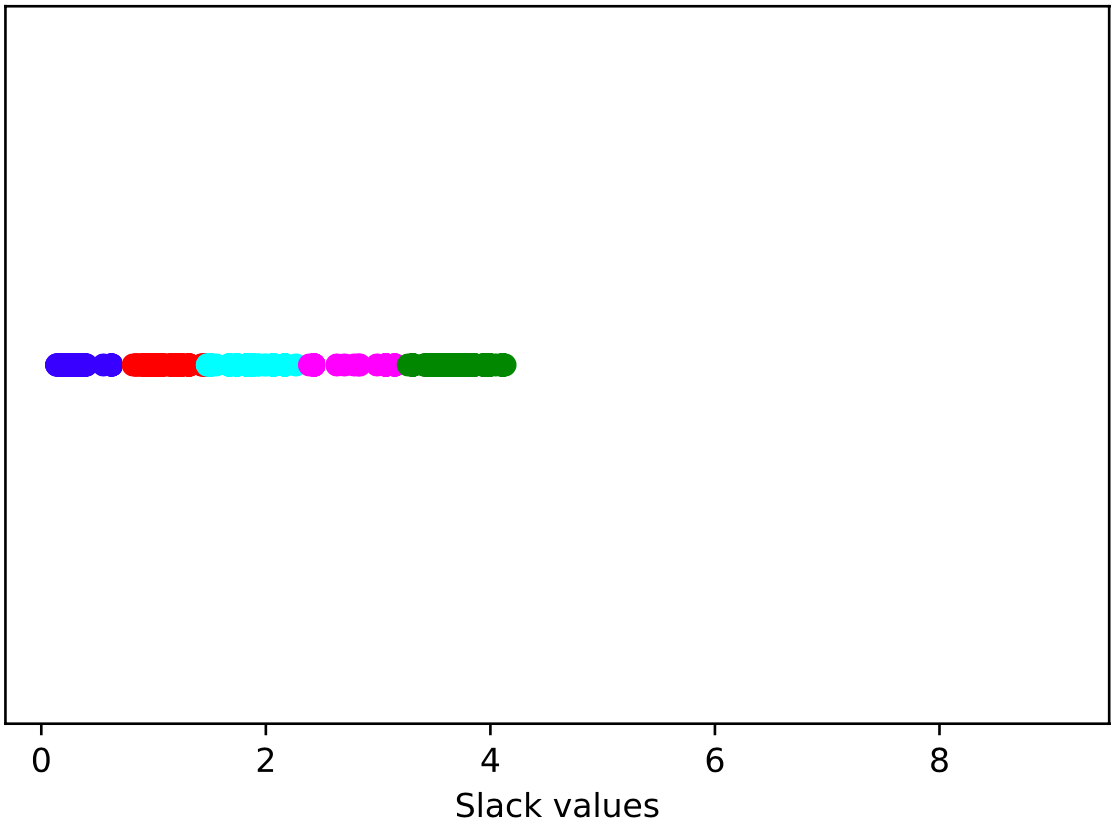}
\caption{Mean-Shift Clustering of Slack of Systolic Array$16\times 16$ }
\label{fig:mshift}
\end{figure}
\end{minipage}
\hspace{1pt}
\begin{minipage}{0.5\textwidth}
\begin{figure}[H]
\centering
\includegraphics[scale=0.17]{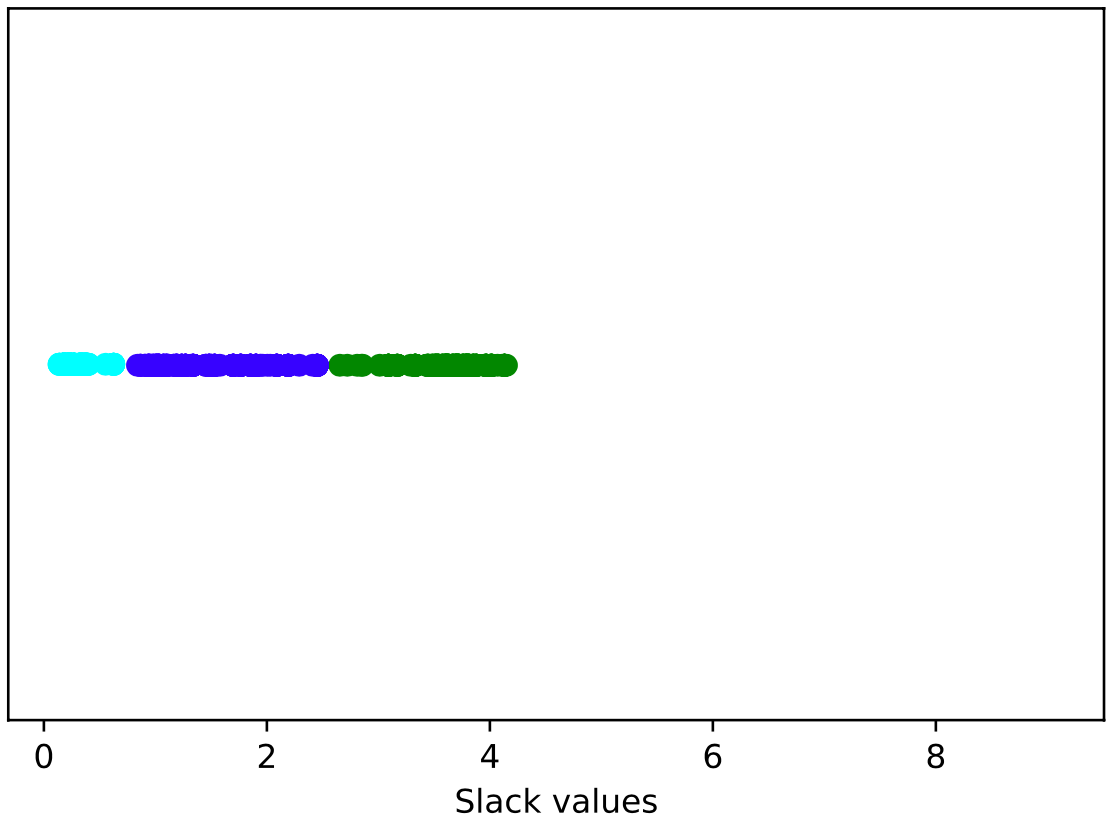}
\caption{DB Scan Clustering of Slack of Systolic Array $16 \times 16$}
\label{fig:dbscan}
\end{figure}
\end{minipage}
\subsection{DBSCAN}
The DBSCAN algorithm has two important hyperparameters, based on which it determines the number of clusters \cite{dbscan}, \textbf{epsilon:} The maximum distance between two samples for one to be considered as in the neighborhood
of the other and \textbf{minpoints:} The number of samples in a neighborhood for a point to be considered as a core point.
 At each step, a data-point that has not been visited before is taken. If there are more data-points than $minpoints$ within its $epsilon$ radius, all the data-points are marked as belonging to a cluster, otherwise 
the first point is marked as noise. For all points in the newly-formed cluster, points within their ‘epsilon’ neighborhood are
checked and labeled as either belonging to a cluster or noise. The process is continued until all data-points have been labeled. 
The greatest advantage of DBSCAN is that it can identify outliers as noise, unlike other algorithms which throw all points into a cluster even if one data point is significantly different from the rest. The time complexity of this algorithm is $O(n)$ for reasonable $epsilon$. This algorithm is not effective for clusters with varying density since $epsilon$ and $minpoints$ are different for different clusters.

\section{Implementation and Result}\label{sec:impl}
As mentioned in Sec. \ref{sec:flow}, the two proposed tool flow has 2 environments. The clustering algorithms for both $Vivado$ and VTR are implemented in Python using the Scikit-learn library. The $synthesis$, $implementation$ and $bit~file~generation$  of $Vivado$ flow is done by the board support package of $Artix-7$ FPGA. The $synthesis$ and $implementation$  of $VTR$ flow is done by the board support package of 22nm, 45nm and 130nm academic FPGAs. As shown in  and Fig. \ref{fig:flow}, the cluster algorithm generates $P$ no. of partition, and the dimension of each partition $(n \times m)$. The static scheme  generates biasing voltages:
\begin{equation}
P \times (n \times m ) \{ V_{ccint_1}, V_{ccint_2},  ... , V_{ccint_i}, ..., V_{ccint_p}\}
\end{equation} 
for $P$ no.of partitions. The runtime scheme calibrate the biasing voltage according to the timing failure detected by Razor placed in every MAC. Runtime scheme gives the final set of biasing voltages: 
\begin{equation}
\begin{split}
V_{ccint_1}+C1.V_s, V_{ccint_2}+C2.V_s,  ... , V_{ccint_i}+Ci.V_s, ...,\\
V_{ccint_p}+ Cp.V_s   
\end{split}
\end{equation} 
Here $C_1$, $C_2$, .. $C_p$ are integers starts from $0$ to any positive value
\begin{table*}[!htb]
	\setlength{\tabcolsep}{2.5pt}
	\caption{Comparison of Dynamic Power (mw) for $Vivado$ and $VTR$ flow}
	\label{res}
	\centering
	\resizebox{14cm}{!}{
		\begin{tabular}{|c|c|c|c|c|c|c|c|}\hline
\shortstack{Schemes under \\25 $^\circ$  Ambient\\ Temperature\\ \&100MHz Clock}&  \shortstack{Dimension of\\ Systolic Array}& \shortstack{Partition\\ No.}& \shortstack{$V_{ccint_i}$\\volt} &   \shortstack{ Vivado\\28nm \\Artix-7}&\shortstack{VTR\\ 22nm} &\shortstack{VTR \\45nm} &\shortstack{VTR\\130nm} \\\hline
\shortstack{Without \\ Voltage\\ Scaling }& \shortstack{ $16 \times 16$\\~\\~} &\shortstack{NA\\~\\~}& \shortstack{1.00\\~\\~}&   \shortstack{408\\~\\~}&269&387&1543\\\hline
&  $8 \times 8$ &\shortstack{partition-1}& 0.96  &&&&\\
Voltage &  $8 \times 8$ &\shortstack{partition-2}& 0.97& 382 &263&380&1531\\
Scaled&  $8 \times 8$ &\shortstack{partition-3}& 0.98& &&&\\
&  $8 \times 8$ &\shortstack{partition-4}& 0.99& &&&\\\hline
\multicolumn{4}{|c|}{\textbf{\% of Reduction}} &\multicolumn{1}{c|}{\textbf{6.37}}&\textbf{1.86}&\textbf{1.8}&\textbf{0.7}\\\hline

\shortstack{Without \\ Voltage\\ Scaling }& \shortstack{ $32 \times 32$\\~\\~} &\shortstack{NA\\~\\~}& \shortstack{1.00\\~\\~}&  \shortstack{1538\\~\\~}&1072&1549&6172\\\hline
&  $16 \times 16$ &\shortstack{partition-1}& 0.96& &&&\\
Voltage &  $16 \times 16$ &\shortstack{partition-2}& 0.97&  1404 &1051&1520&6125\\
Scaled&  $16 \times 16$ &\shortstack{partition-3}& 0.98&  &&&\\
&  $16 \times 16$ &\shortstack{partition-4}& 0.99& &&&\\\hline
\multicolumn{4}{|c|}{\textbf{\% of Reduction}} &\multicolumn{1}{c|}{\textbf{6.76}}&\textbf{1.95}&\textbf{1.87}&\textbf{0.76}\\\hline
\shortstack{Without \\ Voltage\\ Scaling }& \shortstack{ $64 \times 64$\\~\\~} &\shortstack{NA\\~\\~}& \shortstack{1.00\\~\\~}&   \shortstack{5920\\~\\~}&4284&6200&24693\\\hline
&  $32 \times 32$ &\shortstack{partition-1}& 0.96&  &&&\\
Voltage &  $32 \times 32$ &\shortstack{partition-2}& 0.97& 5534&4205&6090&24503\\
Scaled&  $32 \times 32$ &\shortstack{partition-3}& 0.98&  &&&\\
&  $32 \times 32$ &\shortstack{partition-4}& 0.99& &&&\\\hline
\multicolumn{4}{|c|}{\textbf{\% of Reduction}} &\multicolumn{1}{c|}{\textbf{6.52}}&\textbf{1.84}&\textbf{1.77}&\textbf{0.77}\\\hline

\shortstack{Without \\ Voltage\\ Scaling }& \shortstack{ $64 \times 64$\\~\\~} &\shortstack{NA\\~\\~}& \shortstack{0.9\\~\\~}&   \shortstack{not\\supported}&3965&5798&23961\\\hline
&  $32 \times 32$ &\shortstack{partition-1}& 0.7&  &&&\\
Voltage &  $32 \times 32$ &\shortstack{partition-2}& 0.8& not &3818&5656&23631\\
Scaled&  $32 \times 32$ &\shortstack{partition-3}& 0.9&  supported&&&\\
&  $32 \times 32$ &\shortstack{partition-4}& 1.00& &&&\\\hline
\multicolumn{4}{|c|}{\textbf{\% of Reduction}} &\multicolumn{1}{c|}{\textbf{-}}&\textbf{3.7}&\textbf{2.4}&\textbf{1.37}\\\hline
		\end{tabular}
	}
\end{table*}
\subsection{Implementational Challenges}
The proposed design could not be implemented as  none of the present-day FPGA devices support variable voltage scaling in the different logic partitions. The implementation issues of power distribution unit with multiple  $V_{ccint}$ in different partitions are beyond the scope of our paper. However, we consider, the implementation of voltage scaling  technology in ASIC \cite{greentpu} establishes the feasibility of implementation of voltage scaling  technology in FPGA.  
\begin{figure*}[!htb]
\centering
\includegraphics[scale=0.75]{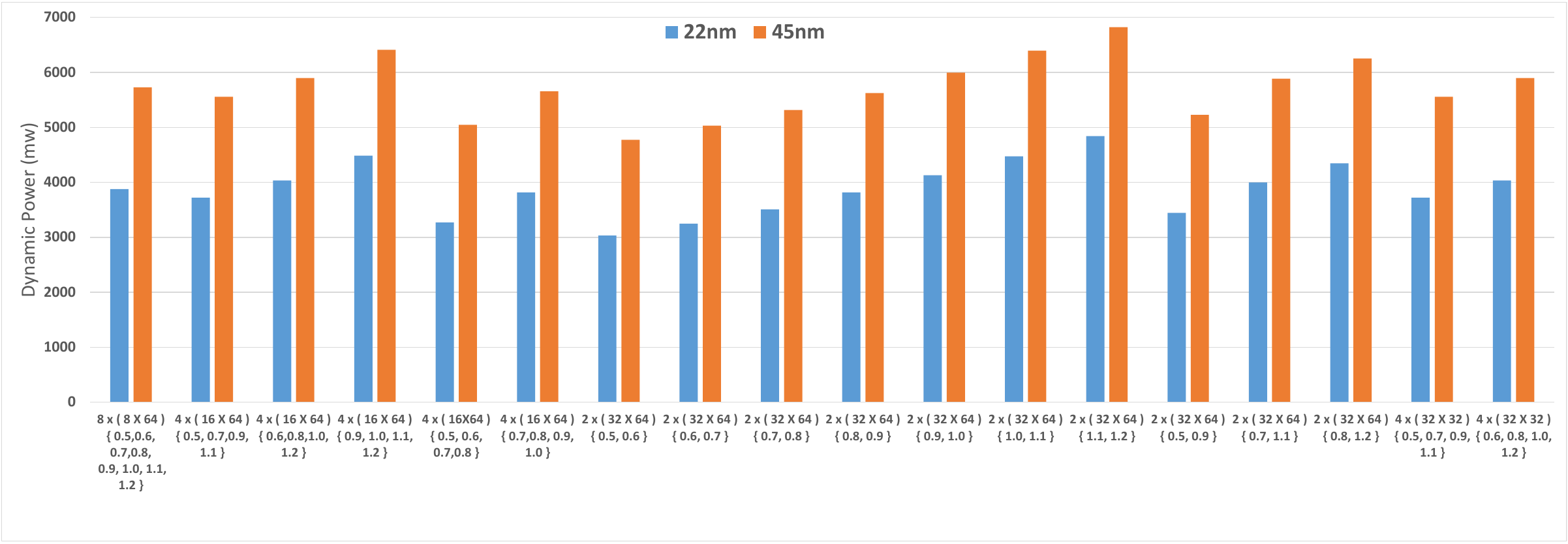}
\vspace{-5pt}
\caption{Comparison of Dynamic Power (mw) of Various Variance of 64 $\times$ 64 Systolic Array on 22nm and 45nm }
\label{fig:22-45}
\end{figure*} 
\begin{figure*}[!htb]
\centering
\includegraphics[scale=0.75]{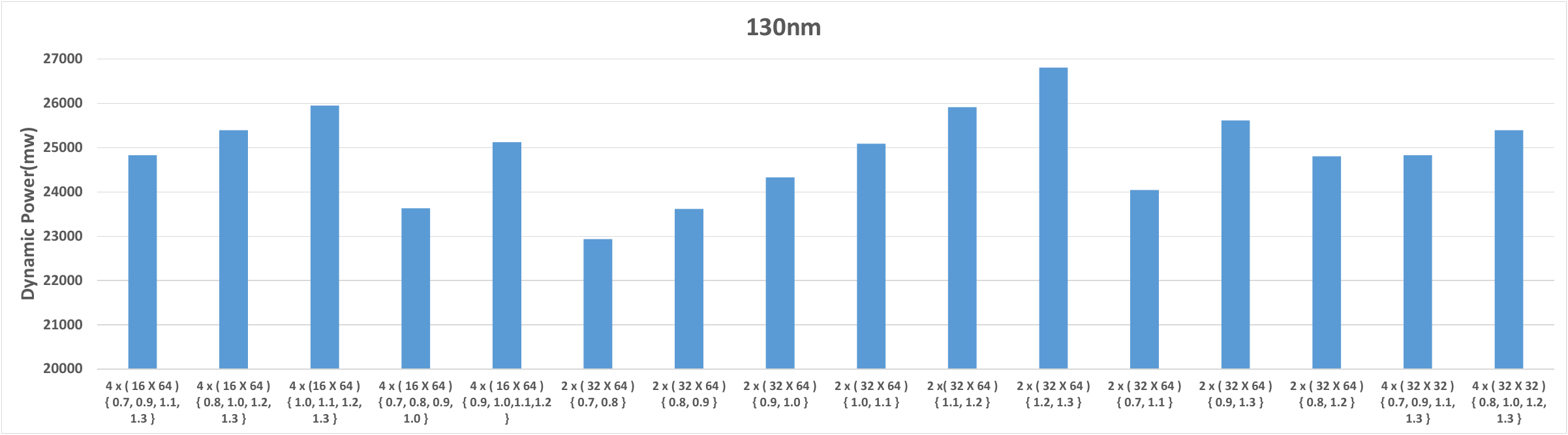}
\vspace{-5pt}
\caption{Comparison of Dynamic Power (mw) of Various Variance of 64 $\times$ 64 Systolic Array on 130nm }
\label{fig:130}
\end{figure*} 
\subsection{Our Validation Strategy}
To validate the claim of the proposal, we have simulated the proposed scheme using VTR and Vivado flow. We have designed a  3 Systolic arrays with the dimension of $16 \times 16$, $32 \times 32$,  and $64 \times 64$. Let us take the example of $16 \times 16$ systolic array where $16 \times 16=256$ MACs are placed in the FPGA. As shown in Fig. \ref{fig:kmean}(b), the K means clustering algorithm mentioned in Sec. \ref{sec:cluster}  divides $16 \times 16$ systolic array into 4 partitions: $partition-1$, $partition-2$, $partition-3$ and $partition-4$. Though the size of the partitions in Fig. \ref{fig:kmean} are not same, for sake of simplicity of implementation we have assumed the same partition size $(8 \times 8)$.  As the current $Vivado$ tool does not allow simulating the design in critical voltage region, our $16 \times 16$ systolic array is tested in the guardband region. Due to the unavailability of multiple $V_{ccint}$ support in single a FPGA dice, our design has been implemented in one partition at a time. Therefore, the power measurement of 4 partitions is also done separately where each partition is considered as an individual circuit. Though $VTR$ allows design in critical regions, for sake of better comparative study we have also used the same voltage ranges used in $Vivado$. 

\subsection{Results}
The guardband region for $Artix-7$ FPGA is 0.95 volt to 1.00 volt. For this example $n=4$, $V_{min}=V_{nom}=1.00~volt$, $V_{crash}=V_{min}=0.95~volt$, therefor $V_b=0.05~volt$. Algorithm \ref{alg:vs}  calculates the $V_{ccint_i}$ of the 4 FPGA partitions of this design which are : $V_{ccint_1}=0.956$ for $partition-1$, $V_{ccint_2}=0.968$ for $partition-2$, $V_{ccint_3}=0.985$ for $partition-3$ and $V_{ccint_4}=0.993$ for $partition-4$. 
We have observed that when the partial sums are moved to the bottom rows of systolic array, the timing error increases significantly \cite{greentpu}. 
In this example the MACs of bottom rows have less minimum slacks, which should be placed in $partition-3$ and $partition-4$ where $V_{ccint_i}$ is more compared to the existing $V_{ccint_i}$. The MACs of upper rows should have more minimum slacks which should be placed in $partition-1$ and $partition-2$ where $V_{ccint_i}$ is less compared to the existing $V_{ccint_i}$. As shown in Fig \ref{fig:part}, the clustering algorithm divides the $16 \times 16$ systolic array  into four $8 \times 8$ systolic array partitions and each partition has $8 \times 8=64$ MACs. The top-left partition-1 consists of a $8 \times 8$ systolic array which has $V_{ccint_1}=0.956\approx 0.96 $. Similarly, top-right partition-2 has $V_{ccint_2}=0.968\approx 0.97$, bottom-left partition-3 has $V_{ccint_3}=0.985\approx 0.98$ and bottom right partition-4 has $V_{ccint_4}=0.993 \approx 0.99$. Table \ref{res} shows the dynamic power consumption of $16 \times 16$, $32 \times 32$ and $64 \times 64$ systolic arrays with 4 partitions. Table \ref{res} shows the adoption of voltage scaling technology reduces 6.97\% to 6.76\% dynamic power consumption for $Vivado$ commercial FPGA  and 0.7\% to 1.95\% VTR academic FPGA. Due to the limited range of voltage in Vivado, these are the lower bounds on improvements. As we can drop biasing voltage closer to NTC for next generation FPGAs, we expect the improvement to improve substantially. In the $4^{th}$ instant, VTR allows more lower biasing voltages which reduces the dynamic power consumption 3.7\%, 2.4\% and 1.37\% for 22nm, 45nm and 130nm respectively.
\par In Fig. \ref{fig:22-45} and Fig. \ref{fig:130}, we have shown dynamic power consumption of different variance of 64 $\times$ 64 systolic array on 22nm, 45nm and 130nm academic FPGAs using VTR flow. Fig. \ref{fig:22-45} and Fig. \ref{fig:130} show that variation of 3 parameters such as number of partition $P$, biasing voltage $V_{ccint_i}$ of each partition and dimension of each FPGA partition $n \times m$  change the dynamic power consumption of 64 $\times$ 64 systolic array by 18\%, 21\% and 39\% for 22nm, 45nm and 130nm academic FPGAs respectively. Here the number of cluster or partition $P$ and dimension of each partition $n \times m$  will be calculated by cluster algorithms. The biasing voltage $V_{ccint_i}$ of each FPGA partition are roughly calculated by static scheme and further calibration of accurate $V_{ccint_i}$ is done by the runtime scheme. Such significant effect of  $P$, $n \times m$ and $V_{ccint_i}$ on dynamic power consumption shows cluster algorithm, static and runtime schemes are very crucial steps of proposed frame. As an example, in Fig. \ref{fig:22-45} and \ref{fig:130}, a name of one variance of systolic array is $4 \times ( 32 \times 32 )
\{ 0.8, 1.0, 1.2, 1.3 \}$ (the rightmost bar at Fig. \ref{fig:130}), here $P$=4, $n \times m$ =$32 \times 32$ and biasing voltages are of 4 partitions are 0.8 volt, 1.0 volt, 1.2 volt and 1.3 volt. In Fig. \ref{fig:22-45} and \ref{fig:130}, the $V_{ccint_i}$ for 130 nm varies from threshold voltage 0.7 volt to 1.3 volt whereas for 22 nm and 45 nm $V_{ccint_i}$ varies from 0.5 volt to 1.2 volt. Though the threshold voltage of 45nm is 0.5 volt and for 22nm it is 0.45 volt, for comparative purposes, in both cases we have measured it form 0.5 volt. It is known that the dynamic power reduces by the square of the supply voltage $V_{ccint_i}$. Also, the  $2 \times ( 32 \times 64 ) \{ 0.5, 0.6 \}$ variant of systolic array implemented in 22nm and 45nm technology have maximum number of MACs which are running with minimum $V_{ccint_i}$  as compared to other reported variants as shown in Fig. \ref{fig:22-45}. Thus, the aforementioned variant consumes minimum dynamic power as compared to other variants reported in Fig. \ref{fig:22-45} .  Going by the same reasoning, the $2 \times ( 32 \times 64 ) \{ 0.7, 0.8 \}$ variant in 135 technology consumes minimum dynamic power when compared to the other variants as reported in Fig \ref{fig:130}. The minimum voltage step of  the power supply \cite{boost} is considered as 0.1 volt. We observe that the timing reports of 16 $\times$ 16, 32 $\times$ 32 and 64 $\times$ 64 systolic arrays before partitioning and after partitioning, which shows very insignificant effects on delay in wires and placement and routing difficulties. Hence, the re-clustering process is not required for the aforementioned systolic arrays.

\section{Conclusion}\label{sec:con}
FPGA is becoming popular in DNN based configurable cloud because of its efficiency and manageability but immoderate power consumption is a growing concern for present FPGA technology. A lot of effort has been made to reduce the power consumption by using multiple biasing voltages in FPGA. This paper proposes a systolic array where the MACs are placed in different partitions of FPGA based on the minimum slacks of different  MACs. Each partition of the FPGA uses different biasing voltage $V_{ccint}$. The proposed $runtime$ and $static$ schemes can tune appropriate $V_{ccint}$ with the group  MACs having similar minimum slacks placed in the partitions. The experimental results show that the voltage scaled systolic array can reduce power consumption. In future we will address two points such as \textbf{(i)} Improvement of $V_{ccint}$ calibration by grouping input sequences with similar delay characteristics to predict future timing failures.  \textbf{(ii)} Study the tradeoff between the DNN accuracy estimated in terms of timing failures with the no. of partitions and that between no. of partitions and dynamic power. The same partition based voltage scaling can be used for other high performance hardware accelerator to reduce the power consumption.

%
\IEEEpeerreviewmaketitle

\bibliographystyle{unsrt}  
\bibliography{IEEEexample}

\end{document}